\documentclass[aps,prl,superscriptaddress,twocolumn,floatfix,a4paper]{revtex4}

\usepackage{graphicx,graphics,epsfig}   
\usepackage{dcolumn}    
\usepackage{bm}         
\usepackage{amsmath}    
\usepackage{verbatim}   
\usepackage{color}      
\usepackage{subfigure}  
\usepackage{times,natbib}
\usepackage{amsmath,amsfonts,amssymb,graphics,graphics,color,times}

\usepackage{latexsym}
\usepackage{amsmath}
\usepackage{amssymb}
\usepackage{amsfonts}
\usepackage{amsthm}
\usepackage{mathrsfs}
\usepackage{color,verbatim,graphics}
\usepackage{psfrag}
\DeclareMathAlphabet{\mathrsfs}{U}{rsfs}{m}{n}
\DeclareMathAlphabet{\mathpzc}{OT1}{pzc}{m}{it}
\DeclareMathAlphabet{\matheus}{U}{eus}{m}{n}
\DeclareMathAlphabet{\mathbbold}{U}{bbold}{m}{n}

\def\one{\leavevmode\hbox{\small1\normalsize\kern-.33em1}}

\newcommand{\ba}{\begin{eqnarray}}
\newcommand{\ea}{\end{eqnarray}}
\newcommand{\ban}{\begin{eqnarray*}}
\newcommand{\ean}{\end{eqnarray*}}

\begin{document}

\title{Closing the detection loophole in multipartite Bell tests using GHZ states
}

\author{K\'aroly F. P\'al}
\affiliation{Institute of Nuclear Research of the Hungarian
Academy of Sciences H-4001 Debrecen, P.O. Box 51, Hungary}

\author{Tam\'as V\'ertesi}
\affiliation{Institute of Nuclear Research of the Hungarian
Academy of Sciences H-4001 Debrecen, P.O. Box 51, Hungary}

\author{Nicolas Brunner}
\affiliation{D\'epartement de Physique Th\'eorique, Universit\'e de Gen\`eve, 1211 Gen\`eve, Switzerland}
\affiliation{H.H. Wills Physics Laboratory, University of Bristol,
Tyndall Avenue, Bristol, BS8 1TL, United Kingdom}

\date{\today}


\begin{abstract}
We investigate the problem of closing the detection loophole in
multipartite Bell tests, and show that the required detection
efficiencies can be significantly lowered compared to the
bipartite case. In particular, we present Bell tests based on
$n$-qubit Greenberger-Horne-Zeilinger states, which can tolerate
efficiencies as low as $38\%$ for a reasonable number of parties
and measurements. Even in the presence of a significant amount of
noise, efficiencies below $50\%$ can be tolerated, which is
encouraging given recent experimental progress. Finally we give
strong evidence that, for a sufficiently large number of parties
and measurements, arbitrarily small efficiencies can be tolerated,
even in the presence of an arbitrary large amount of noise.
\end{abstract}

\maketitle

Quantum nonlocality is arguably one of the most counter-intuitive
aspects of quantum mechanics. According to quantum theory,
separated parties sharing an entangled state and performing
suitably chosen measurements are able to generate
correlations which are unexplainable by any classical mechanism. These nonlocal correlations can be tested
experimentally using Bell inequalities \cite{Bell}. Numerous experiments have demonstrated Bell inequality violations
giving strong evidence that nature is inherently nonlocal
\cite{aspect}. However, technical imperfections in these
experiments open various loopholes, which make it still possible
to explain the data with a local model.
Given the fundamental importance of nonlocality, it is highly
desirable to perform a loophole-free Bell test, which, despite
recent theoretical proposals (see e.g. \cite{CS}) and experimental
progress \cite{harald,steering}, is still missing.

A loophole-free Bell test requires (i) a space-like separation
between the parties, and (ii) a detection efficiency above a
certain threshold (usually high). The first condition ensures that
no communication between the parties is possible, hence closing
the locality loophole. This was achieved in photonic experiments
\cite{aspect}. The second condition ensures that no classical
model exploiting undetected events can reproduce the observed
data, hence closing the detection loophole \cite{pearle}. This was
achieved in atomic experiments \cite{rowe}. However, no
experiments could yet close both loopholes simultaneously. On the
one hand, atomic experiments are unsatisfactory from the locality
point of view. On the other hand, typical photo-detection
efficiencies are still too low to close the detection loophole.

Addressing the detection loophole is also crucial for
information-theoretic applications based on quantum nonlocality
\cite{cc,key,randomness}. Failure in closing the detection
loophole renders these protocols insecure as the observed Bell
violation may have been produced by classical means, as nicely
illustrated by recent experiments faking Bell violations
\cite{fake}.

In general, the required detection efficiency $\eta$ depends on
the Bell inequality and the quantum state which are considered.
For the Clauser-Horne-Shimony-Holt (CHSH) inequality, an
efficiency $\eta>82.8\%$ is required for a maximally entangled
qubit pair, while $\eta>66.7\%$ for a partially entangled state
\cite{Eberhard}. More recently, improvements were reported using
4-dimensional quantum systems, tolerating efficiencies $ \sim
61\%$ \cite{VPB}.  However, from a practical point of view, these
results should be considered carefully, in particular when taking
into account additional imperfections such as background noise.
Importantly, even a small amount of noise increases significantly the threshold efficiencies; in the CHSH
case for instance, adding $1\%$ of noise to the state increases
the threshold from $66.7\%$ to $80\%$ \cite{Eberhard}.

Another approach, which has received so far only little attention,
is to consider multipartite Bell tests, i.e. with $n>2$ observers.
Based on a combinatorial study, Buhrman \emph{et al.} \cite{BHMR}
showed that an arbitrarily small efficiency $\eta$ can be
tolerated as $n$ becomes large. More recently, threshold
efficiencies for the Mermin inequalities were shown to approach
$\eta=50\%$ for large $n$ \cite{CRV}, but remain above $60\%$ for
any practical scenario. The same limit can be approached
for the many-site generalization of the Clauser-Horne inequality
\cite{LS}. Also, a multipartite Bell test based on single-photon
entanglement was shown to approach $\eta=66.7\%$ for large $n$
\cite{CB}. However, up until now, no practical Bell test featuring
efficiencies lower than $60\%$ for all observers was known
\cite{footnote1}.

Here we show that detection efficiencies as low as $38\%$ can be
tolerated in multipartite Bell tests featuring a reasonable number
of parties and measurements, and lower than $50\%$ even in the
presence of noise. Specifically, we present a family of Bell
tests, based on novel Bell inequalities, in which $n$ observers
perform $m$ binary measurements on an $n$-qubit
Greenberger-Horne-Zeilinger (GHZ) state \cite{GHZ}. Notably,
efficiencies $\eta<50\%$ can be tolerated already for a
6-qubit GHZ state and $m=7$ or alternatively for a 5-qubit GHZ
state and $m=11$. Furthermore, the measurements to be performed
are equally distributed on an equator of the Bloch sphere, which is convenient from a practical point of view.
Moreover our Bell tests appear to be robust to noise. For instance for an 8-qubit GHZ state with $10\%$ of noise, efficiencies $\eta\sim50\%$ can be tolerated for $m=7$.
From an experimental perspective, these results look encouraging, given recent experimental progress \cite{exp}, in particular the observation of 8-qubit GHZ states \cite{pan8}.
Finally, we investigate the efficiency for our Bell tests in the asymptotic limit. We give strong evidence that $\eta \rightarrow 2/n$ when $m \rightarrow \infty$, for a pure GHZ state. Moreover we give evidence that arbitrarily low efficiencies can be tolerated, even if an arbitrary amount of noise is added to the GHZ state.

\section{Setup} 

We consider a Bell scenario with $n$ distant observers.
Each observer may choose
between a set of $m$ measurements
$\{A_i\}$, $\{B_j\}$,
$\{C_k\}$ and so on, with $i,j,k,...=0,...,m-1$. All
measurements have binary outcomes, $+1$ and $-1$. We use
the shorthand notation $P(A_iB_jC_k...)\equiv
P(111...|A_iB_jC_k...)$ and similarly for any subset of parties.
We start by defining a family of Bell inequalities:
\begin{align}
&\sum_{i,j,k,\dots=0}^{m-1}P(A_iB_jC_k\dots)(y-x\delta^0_{(i+j+k+\dots) \text{ mod } m})-\nonumber\\
&\sum_{j,k,\dots=0}^{m-1}P(B_jC_k\dots)-\sum_{i,k,\dots=0}^{m-1}P(A_iC_k\dots)-\nonumber\\
&\sum_{i,j,\dots=0}^{m-1}P(A_iB_j\dots)-\dots\leq 0,
\label{eq:Bellineq}
\end{align}
where $\delta^0_{x \text{ mod }m}=1$ if $x$ is divisible by $m$, and is 0 otherwise.
Note that the real parameters $x$ and $y$ are chosen such that the local bound of the Bell inequality is 0. We shall see later how this condition can be enforced.

The observers share a noisy $n$-qubit GHZ state
\begin{equation}
\hat\rho=v|\text{GHZ}\rangle\langle \text{GHZ}|+(1-v)\frac{\one}{2^n},
\label{eq:noisyGHZ}
\end{equation}
with $|\text{GHZ}\rangle\equiv\left(|0\rangle^{\otimes n}+|1\rangle^{\otimes n}\right)/\sqrt 2$
and $v$ is the visibility. This state is fully separable
iff $v\le 1/(1+2^{n-1})$ \cite{DC} and violates a two-setting
full-correlation inequality for $v>1/2^{(n-1)/2}$ \cite{WW}.

Here we will focus on (projective) equatorial qubit measurements, of the form
\begin{align}
\hat A_i&=\cos\varphi^A_i\hat\sigma_x+\sin\varphi^A_i\hat\sigma_y\nonumber\\
\hat B_j&=\cos\varphi^B_j\hat\sigma_x+\sin\varphi^B_j\hat\sigma_y\\
\hat C_k&=\cos\varphi^C_k\hat\sigma_x+\sin\varphi^C_k\hat\sigma_y, \nonumber
\label{eq:measoper}
\end{align}
and so on for all parties; $\hat{\sigma}_{x,y}$ denote the Pauli matrices.

With this choice of measurements and the state \eqref{eq:noisyGHZ}, it follows that (see e.g. \cite{PV11} for details)
\begin{equation}
P(A_iB_jC_k\dots)=
\frac{1+v\cos(\varphi^A_i+\varphi^B_j+\varphi^C_k+\dots)}{2^n}.
\label{eq:nprobcalcphi}
\end{equation}
Next, let us further simplify the structure of the measurement by choosing the $m$ angles to be
evenly distributed around the equator of the Bloch sphere, such that
$\varphi^A_i=\varphi^B_i=\varphi^C_i=2\pi i/m+\pi/n$. With this
choice, we get
\begin{equation}
P(A_iB_jC_k\dots)=
\frac{1-v\cos\left[\frac{2\pi}{m}(i+j+k+\dots)\right]}{2^n}.
\label{eq:nprobcalcijk}
\end{equation}

Finally, since the GHZ state has no $(n-1)$-subcorrelations for equatorial measurements, it follows that all $(n-1)$-particle joint probabilities involved in our inequality take the value $1/2^{n-1}$, independently of $v$.

\section{Threshold efficiencies for reasonable number of parties and measurements} 

All observers detect their particles with the same limited
efficiency $\eta$. In case of non-detection, they agree to output -1. Hence the measurement outputs are still binary and the Bell inequality \eqref{eq:Bellineq} can be used. However the probabilities must be modified in the following way:
$P(A_iB_jC_k\dots) \rightarrow \eta^n P(A_iB_jC_k\dots)$ for $n$-party joint probabilities and
$P(B_jC_k\dots) \rightarrow \eta^{n-1} P(B_jC_k\dots)$ for $n-1$-party joint probabilities.

Bell inequality \eqref{eq:Bellineq} is now violated whenever
\begin{equation}
\left(y-\frac{(1-v)x}{m}\right)\left(\frac{\eta m}{2}\right)^n-n\left(\frac{\eta
m}{2}\right)^{n-1}> 0.
\label{eq:violcond}
\end{equation}
where we have used the fact that $\sum_{i,j,k,...=0}^{m-1} \cos\left[\frac{2\pi}{m}(i+j+k+\dots)\right]=0$.
From equation \eqref{eq:violcond}, the threshold efficiency is then found to be
\begin{equation}
\eta>\eta^*=\frac{2n}{my-(1-v)x}.
\label{eq:violcondeta}
\end{equation}
Thus, in order to determine $\eta^*$ for any given number of parties $n$ and measurements $m$, we must determine the parameters $x$ and $y$ of the Bell inequality \eqref{eq:Bellineq} such that the local bound is 0.
We shall see that, in general, the values of $x$ and $y$ leading to the lowest value of $\eta^*$ may depend on the visibility $v$ of the state.

We recall first that in order to find the maximal value of a linear Bell polynomial (such as \eqref{eq:Bellineq}) it is sufficient to consider local deterministic strategies. For commodity, we denote by $a_i$, $b_j$, $c_k$ and so on the probabilities of
getting outcome +1 for measurement $A_i$, $B_j$, $C_k$ and so on.
We now impose the following condition:
\begin{align}
\label{eq:classical}
&\sum_{i,j,k,\dots=0}^{m-1} a_ib_jc_k ... (y-x\delta^0_{(i+j+k+\dots)\text{ mod }m})-\\
&\sum_{j,k,\dots=0}^{m-1}b_jc_k...-\sum_{i,k,\dots=0}^{m-1}a_ic_k...
-\sum_{i,j,\dots=0}^{m-1}a_ib_j...- ...\leq 0 \nonumber
\end{align}
for any deterministic model that is for any $a_i,b_j,c_k,... \in\{0,1\}$.
Note first that whenever one (or more) parties outputs -1 for all his measurements (say $a_i=0$ for all $i$), then the above condition is indeed satisfied, since only the second sum may be nonzero.
Hence, we can assume that $\alpha\equiv\sum_{i=0}^{m-1}a_i>0$,
$\beta\equiv\sum_{j=0}^{m-1}b_j>0$, $\gamma\equiv\sum_{i=0}^{m-1}c_k>0$, and so on.
Condition \eqref{eq:classical} can then be rewritten as
\begin{equation}
y\leq p+qx,
\label{eq:xyconstraint}
\end{equation}
where $p=\alpha^{-1}+\beta^{-1}+\gamma^{-1}+...$ and $q= S / (\alpha\beta\gamma...)$ and
\begin{align}
S&\equiv\sum_{i,j,k,\dots=0}^{m-1}a_ib_jc_k\dots\delta^0_{(i+j+k+\dots)\text{ mod }m}.
\label{eq:xyconstraintS}
\end{align}
For each choice of $a_i,b_j,c_k,...$, condition
\eqref{eq:xyconstraint} is a linear constraint between $x$ and
$y$, and hence defines a straight line (with positive or zero
slope) in a plane with coordinates $x$ and $y$. For finite values
of $m$ and $n$, we get a finite set of these lines. To ensure that
the local bound of Bell inequality \eqref{eq:Bellineq} is not
greater than 0, $x$ and $y$ must be chosen such that the point
$(x,y)$ lies below all possible straight lines. For small values
of $m$ and $n$, we could find the complete set of straight lines
by exhaustive search. We observed that, although in general the
number of different lines may be very large, only few of them are
relevant for the present problem (see Appendix for
examples and Ref.~\cite{web}).

\begin{figure}[t!]
\vspace{0cm} \includegraphics[width=\columnwidth]{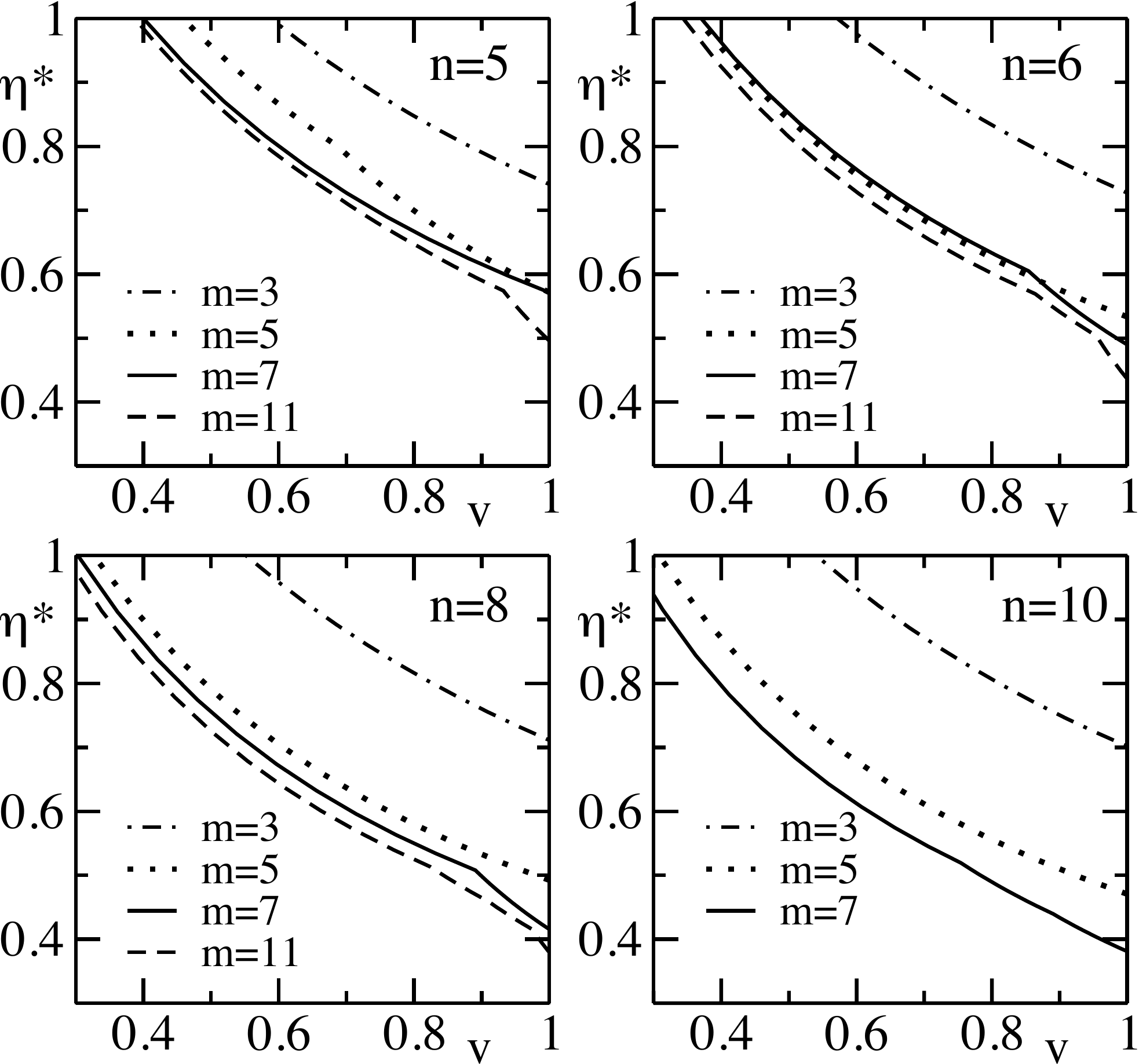}
\caption{Threshold efficiency $\eta^*$ required given an $n$-qubit
GHZ state with visibility $v$. Each observer has $m$ measurement
settings.} \label{fig:det}
\end{figure}

We are now ready to present our main results, namely the threshold
detection efficiencies $\eta^*$ for our Bell tests considering all
number of parties and measurements which maybe reasonable from an
experimental viewpoint. Up to $n=10$ parties and $m=11$
measurements, we could determine, for any value of the visibility
$v$, the optimal Bell inequality (given by $x$ and $y$) leading to
the lowest efficiency $\eta^*$ (see Fig.~1 and \cite{web}).
Notably, efficiencies below $38\%$ can be reached for a $n=8$ and
$m=11$. Also, $\eta^*<50\%$ can be obtained for a 5-qubit GHZ
state with $m=11$ (for $n=8$, $m=5$ is sufficient). Perhaps even
more importantly, these threshold efficiencies appear to be robust
to noise. In most cases, adding $10\%$ noise increases $\eta^*$ by
only a few percents. For instance, for $n=8$ and $m=11$, one has
$\eta^*<50\%$ even for $v$ as low as $85\%$. This shows that
multipartite Bell test can tolerate significantly lower detection
efficiencies, even in the presence of noise, compared to all
bipartite Bell test proposed so far with a reasonable number of
measurements and dimensions. Finally note that we focused here on
cases in which $m$ takes prime values; other cases are much less
favorable.

\section{Asymptotic limit} 

From a theoretical point of view it
is also interesting to investigate the behaviour of $\eta^*$ in
the asymptotic limit, i.e. for $n $ and/or $m$ large.

The main difficulty consists in deriving the optimal parameters
$x$ and $y$ for Bell inequalities \eqref{eq:Bellineq} for
arbitrary $n$ and $m$. Although we have not been able to find a
general solution, we could solve this problem by considering only
a subset of all deterministic strategies which we conjecture to be
optimal. From our investigation for small values of $m$ (prime)
and $n$, we observed that the set of relevant straight lines,
delimiting the region of allowed values of $x$ and $y$, are always
given by deterministic strategies with a simple and regular
structure. Such strategies, which from now on we term 'regular
arrangements', are as follows. For all parties but one (say A),
the output will be +1 for the measurements of lowest
indices and -1 for the remaining ones; more formally, $b_j=1$ iff
$j<\beta$, $c_k=1$ iff $k<\gamma$ and so on. For party A, we
consider strategies of the form $a_i=1$ iff
$i=i_0,...,,(\alpha+i_0)\text{ mod } m $ for $i_0=0,..,m-1$.
Moreover, it turns out that it is enough to consider strategies in
which $\alpha,\beta,\gamma,...$ differ form each other by at most
one.

We start by considering the case of a pure $n$-qubit GHZ state, i.e. $v=1$. In this case, the threshold efficiency depends only on the parameter $y$ (see Eq. \eqref{eq:violcondeta}). As we would like to choose $y$ as large as possible, we are looking for the straight line of the form \eqref{eq:xyconstraint} with zero slope, i.e. with $S=0$.
For regular arrangements, we can derive the optimal efficiencies $\eta^*$ for arbitrary $m$ (prime) and $n$. Note first that, for a regular arrangement to achieve $S=0$, the total number of measurements for which the outcome is +1 must be upper bounded: $\alpha+\beta+\gamma+\dots\leq n+m-2$. To see this, consider first the case $\beta=\gamma=...=1$, i.e. $b_j=c_k=...=1$ iff $j=k=..=0$. To ensure that $S=0$, one must choose $a_0=0$, which leads to  $\alpha\leq m-1$, hence finally to $\alpha+\beta+\gamma+\dots\leq n+m-2$. If we then increase $\beta$ by one, we must now also impose that $a_{m-1}=0$, hence decreasing $\alpha$ by one. Thus the total number of measurements with outcome +1 does not increase.
Next, one has to maximize the value of $p$ (see Eq. \eqref{eq:xyconstraint}). Given that the total number of measurements with outcome +1 is upper bounded, we get the largest value of $p$ by distributing these measurements as evenly as possible between the $n$ observers. Hence we get $\alpha=\beta=\gamma =...= (m+n-2)/n$, leading to $y \leq y_{max}= n^2 / (m+n-2)$; for simplicity we have assumed here that $m-2$ is a multiple of $n$. This leads to the threshold efficiency
\begin{equation}
\eta^* = \frac{2}{n}+\frac{2}{m}-\frac{4}{mn}.
\label{eq:etaminspec}
\end{equation}
Hence Bell inequality \eqref{eq:Bellineq} can be violated using detectors with arbitrarily low efficiency $\eta>0$, by choosing $n$ and $m$ large enough.
Note that if either $n$ or $m$ is finite, $\eta^*$ tends to a strictly positive value. Note also that for any given number of parties $n$, we have that $\eta\rightarrow 2/n$, for sufficiently large $m$. This improves on the results of Ref.~\cite{BHMR}, which had $\eta\rightarrow 8/n$.
Finally, note that we have again considered only $m$ prime. For $m$ not prime, it is possible to have $\alpha+\beta+\gamma+\dots> n+m-2$ such that $S=0$.

\begin{figure}[t]
\vspace{0cm}
\includegraphics[width=0.7\columnwidth]{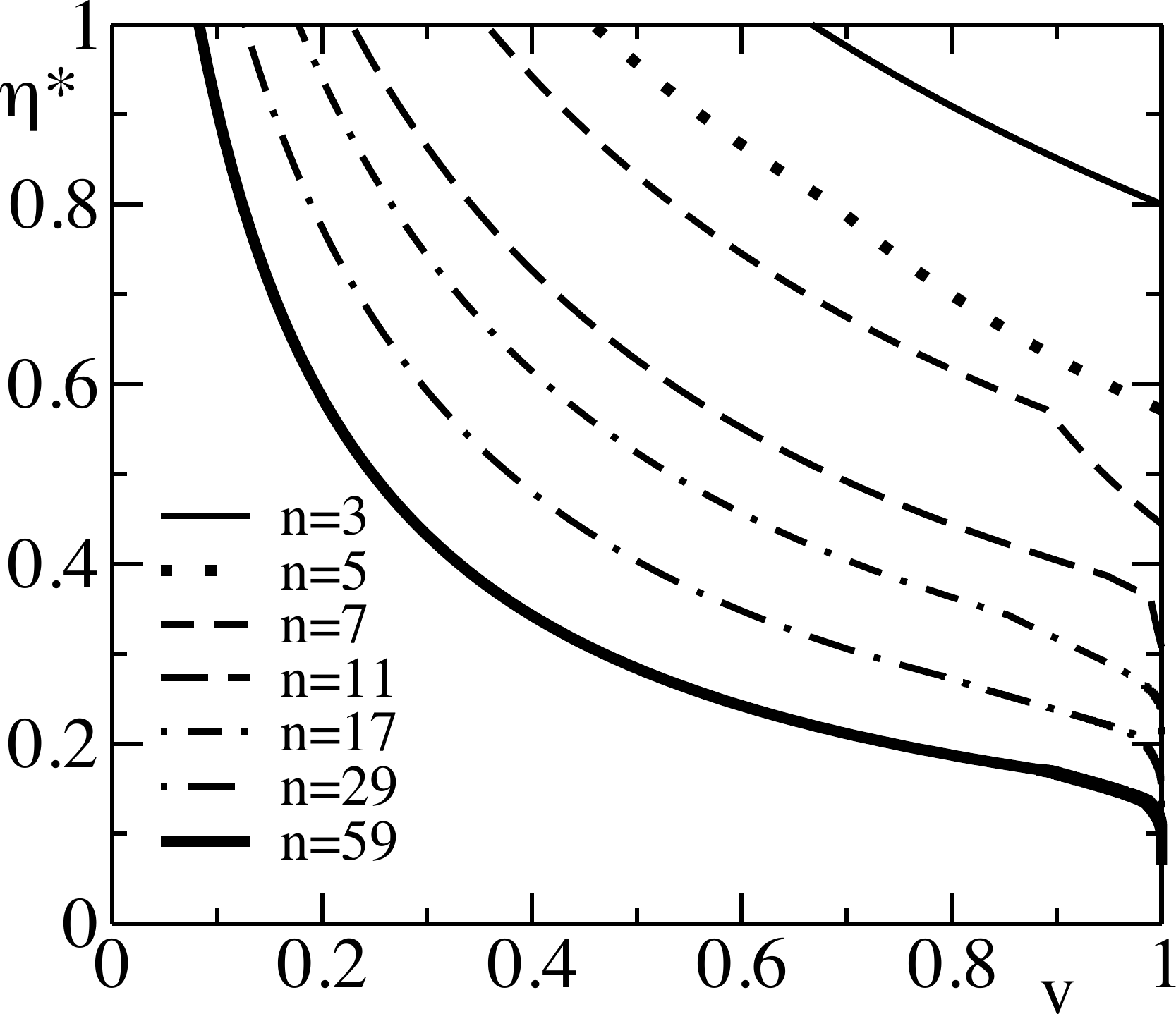} \caption{Threshold detector
efficiency $\eta^*$ versus the visibility $v$ of the GHZ state, in
the case $m=n$. This indicates that in the limit of large $n$ and
$m$, $\eta^*$ can become arbitrarily small, even for small
visibilities $v$.} \label{fig:veta}
\end{figure}

Next, we investigate the case in which the visibility of the state is limited, i.e. $0<v<1$, and give evidence that our Bell tests can tolerate arbitrarily low detection efficiencies even in the presence of an arbitrarily large amount of noise, when taking $n$ and $m$ large enough.
We first determined for $n=m\leq 59$ the optimal Bell inequalities (i.e. parameters $x,y$), assuming that the optimal local deterministic strategy is a regular arrangement. The results, shown in Fig.~2, support qualitatively our above claim.


Then, we consider the case $m \ll n$. As we could not derive the complete set of conditions on $x$ and $y$, we focused our efforts, as in the noiseless case, on the horizontal line, i.e. $y=y_{max}=n+1-m/2$ (for $m \ll n$). From Eq. \eqref{eq:violcondeta}, one can see that $x$ must be taken as small as possible in order to get the lowest values of $\eta^*$. Hence our goal here is to determine the smallest possible value of $x$, i.e. $x_{min}$, such that all conditions \eqref{eq:xyconstraint} hold.
We conjecture that $x_{min} \leq nm$ holds for $m \ll n$, leading to a threshold efficiency
\begin{equation}
\eta^* \simeq \frac{2}{mv}.
\end{equation}
Hence, even in the case of arbitrarily small visibility $v$, $\eta^*$ can become arbitrarily small by taking $m \ll n$ large enough.
To support our conjecture that $x_{min} \leq nm$, we checked that, for $m \leq 199$ (prime) and $n\leq 199$, $x_{min}$ is always achieved by only two possible strategies: (i) all parties output +1 for all measurements, i.e. $\alpha=\beta=\gamma=...=m$, leading to $p=n/m$, $S=m^{n-1}$ and $q=1/m$; this corresponds to a line reaching $y=y_{max}$ at $x_1= m y_{max}-n$; or (ii) a regular arrangement with $\alpha+\beta+\gamma+...=m+n$, leading to $p=n-m/2$, $S=2$, and $q=1/2^{m-1}$; this corresponds to a line reaching $y=y_{max}$ at $x_2 = (y_{max}-p)/q=2^{m-1}$, independent of $n$. Indeed $x_1,x_2\leq nm$ when $m \ll n$.

\section{Conclusion} 

We presented a family of multipartite
Bell tests and derived the minimal detection efficiencies required
in order to close the detection loophole. Notably, efficiencies
below $50\%$ can be tolerated for a reasonable number of parties
and measurements, even in the presence of significant amount of
noise. Our Bell tests are based on $n$-qubit GHZ states, which
have been realized experimentally. In particular, Ref. \cite{pan8}
recently reported 8-qubit GHZ entanglement, with fidelities of
$\sim 70\%$. This would require a detection efficiency of $\sim
60\%$ in our Bell tests, which seems within reach of current
photonic experiments \cite{steering}. However, the main challenge is to achieve a heralded preparation of the GHZ state \cite{walther}. Nevertheless, this shows that the
multipartite setting offers possibilities for a
loophole-free Bell test based on photons. More generally, we
believe that our findings open interesting experimental
perspectives for multipartite nonlocality, and for its
applications \cite{cc,Aolita}.

\emph{Acknowledgements.} The authors thank C. Simon for
discussions, and acknowledge financial support from the UK EPSRC, the Swiss National Science Foundation (grant PP00P2\_138917), the EU DIQIP, the Hungarian National Research Fund OTKA (PD101461), a J\'anos Bolyai Grant of the Hungarian Academy of Sciences, and the
T\'AMOP-4.2.2.C-11/1/KONV-2012-0001 project. The project has also been
supported by the European Union, co-financed by the European
Social Fund.

\section{Appendix}

Here we give more details concerning the choice of parameters $x$
and $y$, defining our Bell inequalities \eqref{eq:Bellineq}, such
that the local bound is 0. As explained in the main text, one must
check a finite set of conditions, of the form
\eqref{eq:xyconstraint}, which define straight lines in the plane
with coordinates $x$ and $y$. For small values of $n$ and $m$ the
complete set of lines can be found. Fig.~3 illustrates the
situation for the case of $n=5$ observers, and up to $m=13$
measurements. Note that although the total number of lines is
large, only few ones turn out to be relevant. Also the optimal
choice of $x$ and $y$, which may depend on the visibility of the
state $v$ (see Eq. \eqref{eq:violcondeta}), is always one of the
intersections of two (or more) lines (marked by dots in Fig.~3).
The (x,y) pairs, along with the ranges of visibilities
where they are optimal choice, are shown in Ref.~\cite{web}. These
values have been used to generate the pieces of the curves shown
in Fig.~1.

\begin{figure}[t!]
\vspace{0cm}
\includegraphics[width=8 cm]{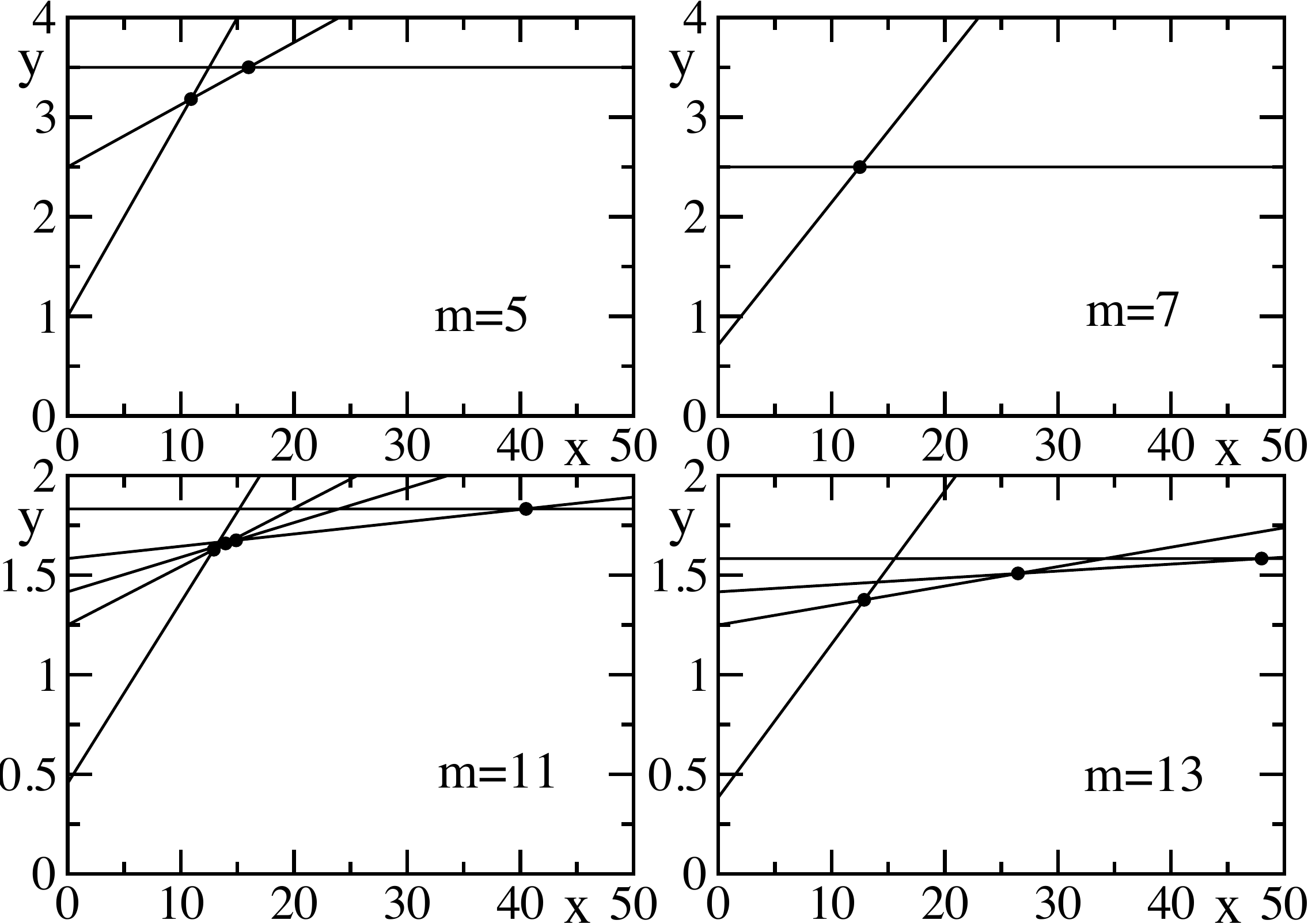} \caption{The set of straight lines
defining the region of allowed parameters $x$ and $y$. Each
straight line corresponds to one condition as given in Eq.
\eqref{eq:xyconstraint} (with equality). Hence all points $(x,y)$
located below all lines will satisfy all conditions
\eqref{eq:xyconstraint}. Hence for any such choice of parameters
$x$ and $y$, the local bound of the Bell inequality will be
smaller or equal to 0 as desired. Here we have $n=5$ parties. Only
the relevant lines are shown.} \label{fig:border5}
\end{figure}


\begin{thebibliography}{99}

\bibitem{Bell} J. S. Bell, Physics {\bf 1} 195 (1964).

\bibitem{aspect} A. Aspect, Nature {\bf 398}, 189 (1999).

\bibitem{CS} D. Cavalcanti \emph{et al.}, Phys. Rev. A {\bf 84}, 022105 (2011); A. Cabello and F. Sciarrino, Phys. Rev. X {\bf 2}, 021010 (2012); J.~B.
Brask and R. Chaves, Phys. Rev. A {\bf 86}, 010103 (2012); C. Teo \emph{et al.}, arXiv:1206.0074; Y. Lim \emph{et al.}, Phys. Rev. A {\bf 85}, 062112 (2012).

\bibitem{harald} J. Hofmann \emph{et al.}, Science {\bf 337}, 72 (2012).

\bibitem{steering} B. Wittmann \emph{et al.}, New J. Phys. {\bf 14}, 053030 (2012); D.-H. Smith \emph{et al.}, Nat. Commun. {\bf 3}, 625 (2012);
A.~J. Bennett \emph{et al.}, Phys. Rev. X {\bf 2}, 031003 (2012).

\bibitem{pearle}
P.~M. Pearle, Phys. Rev. D {\bf 2}, 1418 (1970); see also C. Branciard, Phys. Rev. A {\bf 83}, 032123 (2011).

\bibitem{rowe} M.~A. Rowe \emph{et al.}, Nature {\bf 409}, 791 (2001).

\bibitem{cc}
H. Buhrman, R. Cleve, S. Massar, and R. de Wolf, Rev. Mod. Phys.
{\bf 82}, 665 (2010).

\bibitem{key}
A. Acin et al., Phys. Rev. Lett. {\bf 98}, 230501 (2007).

\bibitem{randomness}
S. Pironio et al., Nature (London) {\bf 464}, 1021 (2010); R.
Colbeck, Ph.D. Thesis, University of Cambridge (2006); R. Colbeck
and A. Kent, J. Phys. A: Math. Th. 44, 095305 (2011).

\bibitem{fake} I. Gerhardt \emph{et al.}, Phys. Rev. Lett. {\bf 107}, 170404 (2011);
E. Pomarico \emph{et al.}, New J. Phys. 13, 063031 (2011); D.S. Tasca \emph{et al.}, Phys. Rev. A 80, 030101(R) (2009).


\bibitem{Eberhard}
P.H. Eberhard, Phys. Rev. A {\bf 47}, R747 (1993); see also G. Lima \emph{et al.}, Phys. Rev. A {\bf 85}, 012105 (2012).

\bibitem{VPB}
T. V\'ertesi, S. Pironio, and N. Brunner, Phys. Rev. Lett. {\bf 104}, 060401 (2010).

\bibitem{BHMR}
H. Buhrman, P. H{\o}yer, S. Massar, and H. R\"ohrig, Phys. Rev.
Lett. {\bf 91}, 047903 (2003).

\bibitem{CRV}
A. Cabello, D. Rodriguez, and I. Villanueva, Phys. Rev. Lett. {\bf
101}, 120402 (2008).

\bibitem{LS}
J.-A. Larsson and J. Semitecolos, Phys. Rev. A {\bf 63}, 022117
(2001).

\bibitem{CB}
R. Chaves and J.B. Brask, Phys. Rev. A {\bf 84}, 062110 (2011).


\bibitem{footnote1} Note that in an asymmetric setup, where parties may have different efficiencies (e.g. in atom-photon entanglement), lower efficiencies were reported, see \cite{Asym}.


\bibitem{Asym} N. Brunner, N. Gisin, V. Scarani, and C. Simon, Phys. Rev. Lett. {\bf 98}, 220403 (2007); A. Cabello and J.A. Larsson, Phys. Rev. Lett. {\bf} 98, 220402
(2007).

\bibitem{GHZ}
D.M. Greenberger, M.A. Horne, and A. Zeilinger, {\it Bells
Theorem, Quantum Theory, and Conceptions of the Universe} (ed. M.
Kafatos, Kluwer Academic, Dordrecht, Holland, 1989), pp. 69–-72.

\bibitem{exp} C.-Y. Lu et al., Nat. Phys. {\bf 3}, 91 (2007); R. Prevedel et al., Phys. Rev. Lett. {\bf 103}, 020503 (2009); W. Wieczorek et al., Phys. Rev. Lett. {\bf 103}, 020504 (2009).

\bibitem{pan8} X.-C. Yao \emph{et al.}, Nature Phot. {\bf 6}, 225 (2012).

\bibitem{DC}
W. D\"ur and J.I. Cirac, Phys. Rev. A {\bf 61}, 042314 (2000).

\bibitem{WW}
R.F. Werner and M.M. Wolf, Phys. Rev. A {\bf 64}, 032112 (2001).

\bibitem{PV11}
K.F. P\'al and T. V\'ertesi, Phys. Rev. A {\bf 83}, 062123 (2011).

\bibitem{web} For all data, see: \verb+http://www.atomki.hu/+\\
\verb+atomki/TheorPhys/Deteff_GHZ/BellXY.html+

\bibitem{walther} P. Walther, M. Aspelmeyer, and A. Zeilinger, Phys. Rev. A {\bf 75}, 012313 (2007).

\bibitem{Aolita}
V. Scarani and N. Gisin, Phys. Rev. Lett. {\bf 87}, 117901 (2001); J. Silman \emph{et al.}, Phys. Rev. Lett. {\bf 106}, 220501 (2011);
L. Aolita, R. Gallego, A. Cabello, and A. Acin, Phys. Rev. Lett. {\bf 108}, 100401 (2012).



\end{thebibliography}
\end{document}